\documentclass[aps,amsmath,amssymb,preprint,nofootinbib]{revtex4}
\usepackage{graphicx}
\usepackage{xcolor}
\usepackage{adjustbox}
\usepackage{ulem}
\usepackage{hyperref}
\hypersetup{
    colorlinks=true,
    linkcolor=blue,
    citecolor=blue,
    urlcolor=blue
}

\def\gsim{\mathrel{\rlap{\lower4pt\hbox{\hskip1pt$\sim$}}
		\raise1pt\hbox{$>$}}}       %greater than or approx. symbol

\begin{document}
	
 	\title{Probing Grand Unification with Quantum Sensors}
	\author{Xavier Calmet} \email{x.calmet@sussex.ac.uk}  \author{Nathaniel Sherrill} \email{n.sherrill@sussex.ac.uk}
	\affiliation{Department of Physics and Astronomy,\\
		University of Sussex, Brighton, BN1 9QH, United Kingdom}	

	\begin{abstract}
 We discuss how grand unification can be probed with experiments at low energies using quantum sensors. 	
 Specifically, we show that scalar multiplets coupled to the gauge sector of a grand unified theory provide a mechanism for a time-varying unified coupling which has low-energy consequences which can be probed with quantum sensors. We then assume that the multiplets represent ultra light dark matter. Constraints on ultra light dark matter couplings to regular matter are extracted using atomic clock comparisons, pulsar timing arrays (NANOGrav) and MICROSCOPE.
\end{abstract}
	
	\maketitle
\section{Introduction}

The stability of fundamental constants has long been questioned starting with the seminal work of Dirac~\cite{Dirac}. Since then, this topic has been revisited frequently by famous scientists. E.g., in 1996 Dyson and Damour \cite{Damour:1996zw} studied bounds on the stability of the fine-structure constant using data from Oklo. Some theories of quantum gravity, typically involving extra dimensions, predict a cosmological time evolution of fundamental constants which are then moduli, see e.g., ~\cite{Polchinski:1998rq,Polchinski:1998rr,Marciano84,CO94,Fritzsch:2016ewd}. A comprehensive review of the fundamental constants and their variation can be found in \cite{Uzan:2002vq}. 

Due to recent technological progress, quantum sensors and other atomic clocks (see e.g. \cite{Sherrill:2023zah,Barontini:2021mvu,Filzinger23,Chou:2023hcc,Buchmueller:2022djy}) have made impressive advances which enable to test the stability of fundamental constants with 
much improved precision. These technological advances have revitalised the field and led to new ideas of how to probe fundamental physics using quantum sensors.

Recently, there has been significant interest in fundamental constants which are effectively time dependent due to the presence of time-evolving scalar fields. These scalar fields can be thought of as giving a time-dependent contribution to the fundamental constant. This is the case, for example, with models of ultra light dark matter \cite{DP14,SF14,AHT15}, which have been now fairly extensively studied. 

It has also been pointed out that a time variation of fundamental constants could enable us to probe grand unified theories at very low energy without ever producing any of the heavy particles associated with such models \cite{CF2002}. This is because the unification mechanism imposes relations between the time variation of, e.g., the fine-structure constant and, e.g., the proton mass. To the best of our knowledge, there has not been any concrete proposal to generate a time variation of the unified coupling constant within models of grand unification. 

The aim of this paper is twofold. We first provide a mechanism to explain a time variation of the unified gauge coupling within the context of grand unified theories. Our mechanism involves a scalar multiplet charged under the grand unified theory group. We then assume a special case where the multiplet is identified with ultra light dark matter. Ultra light dark matter is a topic which has received a considerable amount of attention recently (see, e.g., recent reviews~\cite{SMDM}). As far as we are aware, this work is the first attempt to embed this dark-matter candidate in a grand unified theory.

This paper is organized in the following way. We first describe how a scalar multiplet of a grand unified group can generate an effective time variation of the unified coupling constant of this grand unification group via dimension-five operators generated by quantum gravity. We investigate the consequences for low-energy coupling constants. We then assume that the scalar multiplet accounts for dark matter and posit that the scalar multiplet is ultra light. We derive limits on the coupling between the scalar multiplet and standard model particles using data from clocks, NANOGrav and MICROSCOPE. We then conclude.

\section{A mechanism for the time variation of fundamental constants}

It was pointed out a while ago \cite{CF2002} that a time variation of fundamental constants of the standard model would enable us to probe unification physics, i.e., a very high-energy scale of the order of $10^{16}$\;GeV with low-energy experiments involving, e.g., atomic clocks.

An effective time variation of the unified coupling constant can be generated by a dimension-five operator involving scalar multiplets $H$ and the field strength of the unified theory ${\cal G}_{\mu\nu}$. Operators of the type $\text{Tr}(H \cal{G}_{\mu\nu} \cal{G}^{\mu\nu})$ will lead to an effective time variation of the unified coupling constant if $H$ are time-dependent classical fields.

While the mechanism highlighted here is generic and can be embedded in any grand unified theory, for definiteness we consider supersymmetric $SU(5)$ as a unification group. We assume that supersymmetry is broken at some low-energy scale and that unification takes place at some $10^{16}$\;GeV. Note that a time variation of fundamental constants due to a background field has been considered previously in the context of axion physics \cite{HR84,HST90}.

Starting from the standard ${\cal N}=1$ supersymmetric $SU(5)$ grand unified theory {\`a} la Georgi and Glashow,  we consider additional dimension-five effective operators consisting of a unified gauge field strength ${\cal G}^a_{\mu\nu}$ and scalar multiplets $H^{ab}_r$:
\begin{equation} \label{L1a}
    {\cal L} =  -\frac{1}{4}{\cal G}_{\mu\nu}^a {\cal G}^{a\mu\nu} -\frac{1}{4}\sum_{r} g_{u}^2 \frac{c_r}{M_{\rm P}}H^{ab}_r {\cal G}_{\mu\nu}^a {\cal G}^{b\mu\nu},
\end{equation}
where $M_{\rm P}$ is the reduced Planck mass, $g_u$ the unified coupling constant, and $c_r$ are dimensionless coefficients where $r$ denotes the scalar multiplet representation. Quantum gravity will generically generate dimension-five and higher operators (see, e.g., Refs.~\cite{Hill84,SW84,tHooft80}). We restrict ourselves to the leading-order quantum-gravitational corrections to the $SU(5)$ Lagrangian density. As $H^{ab}_r$ are charged under $SU(5)$, the Wilson coefficients depend on the coupling $g_{u}^2$. 

Our aim is to study the time dependence of the standard model within this unified theory framework. The fine-structure constant, Fermi constant and the strong-coupling constant will be time dependent if the unified coupling constant receives time-dependent corrections. The easiest way to achieve this is to assume that scalar multiplets $H^{ab}_r$ are time-dependent classical background fields. They could be space-time dependent, but we limit attention to their time dependence. We will first show that the dimension-five operators lead to an effective time variation of the unified coupling constant. These dimension-five operators thus represent a mechanism for the time variation of the unified coupling constant posited in \cite{CF2002}. 

We will assume that the scalar multiplets $H^{ab}_r$ can be much lighter than the unification scale $M_u$. This does not lead to any issue with typical tests of grand unified theories, such as proton decay, as we do not couple these extra multiplets to fermions. Explaining why scalar fields are lighter than unification scale in grand unified theories (e.g., the mass of the Higgs boson) has been a concern for many authors. Here, we do not attempt to justify the masses of $H^{ab}_r$ which we cannot a priori calculate from first principles. As these parameters need to be renormalized, we do not consider this a mathematically well-posed question within the realm of quantum field theory. 

Rescaling the gauge fields by $g_{u}^{-1}$ makes it easier to identify the effective time-dependent unified coupling constant corresponding to the canonically normalized kinetic term. We obtain
\begin{equation} \label{L1}
    {\cal L} =  -\frac{1}{4 g_{u}^2}{\cal G}_{\mu\nu}^a {\cal G}^{a\mu\nu} -\frac{1}{4}\sum_{r}\frac{c_r}{M_{\rm P}}H^{ab}_r {\cal G}_{\mu\nu}^a {\cal G}^{b\mu\nu},
\end{equation}
and find $({\bar{g}_{u}})^{-2}=(g_{u})^{-2}+\sum_{r}(c_r/M_{\rm P})H^{aa}_r$, where $H^{aa}_r$ is the singlet component of $H_r^{ab}$ in terms of the standard model gauge symmetries. In other words, we now have a unified coupling constant 
\begin{eqnarray}
\alpha_u(t)=\frac{{\bar{g}_{u}^2}}{4\pi},
\end{eqnarray}
which is effectively time dependent due to $H^{aa}_r$.  

After $SU(5)$ symmetry breaking, $H^{aa}_r$  contributes to the gauge kinetic terms such that at the unification scale
\begin{equation} \label{L2}
%\label{Leps}
    {\cal L} =  -\frac{1}{4}\left(\frac{1}{g_{1}^2}+\epsilon_1 \right)F_{\mu\nu} F^{\mu\nu}_{U(1)} -\frac{1}{2}\left(\frac{1}{g_{2}^2}+\epsilon_2\right){\rm Tr}F_{\mu\nu} F^{\mu\nu}_{SU(2)} -\frac{1}{2}\left(\frac{1}{g_{3}^2}+\epsilon_3\right){\rm Tr}F_{\mu\nu} F^{\mu\nu}_{SU(3)},
\end{equation}
where 
\begin{equation}
\label{eps}
\epsilon_i = \sum_r \frac{c_r}{M_{\rm P}}H_r\delta_i^{(r)},
\end{equation}
where we set $H_r^{aa}=H_r\delta_i^{(r)}$. Note that in our notation $g_1$ contains the $SU(5)$ embedding factor $\sqrt{5/3}$ which we will make explicit when needed.
The time-dependent couplings for $U(1)_Y, SU(2)_L$, and $SU(3)_c$ are given in terms of the time-independent couplings $g_i$ and the $\epsilon_i$ by
\begin{equation}
\label{timedepcou}
(\bar g_{i})^{-2}=(g_{i})^{-2}+\epsilon_i.
\end{equation}
The Clebsch-Gordan coefficients $\delta_i^{(r)}$ for the gauge-invariant irreducible representations are given in Table~\ref{table:1}~~\cite{slansky,CHR}. For $SU(5)$, the irreducible representation $r =$ \textbf{1, 24, 75, 200} contain exactly one standard model singlet. For other groups, e.g., $SO(10)$, there may be more than one standard model singlet. 
\begin{table}[h]
\caption{Clebsch-Gordan coefficients in Eq.~\eqref{eps}.}
\begin{center}
\begin{tabular}{ c c c c} 
 \hline\hline
 $SU(5)$ irrep. $r$ & $\delta_1^{(r)}$ & $\delta_2^{(r)}$ & $\delta_3^{(r)}$\\
 \hline
 \textbf{1} & $-1/\sqrt{24}$ & $-1/\sqrt{24}$ & $-1/\sqrt{24}$ \\ 
\textbf{24} & $1/\sqrt{63}$ & $3/\sqrt{63}$ & $-2/\sqrt{63}$ \\ 
\textbf{75} & $5/\sqrt{72}$ & $-3/\sqrt{72}$ & $-1/\sqrt{72}$ \\ 
\textbf{200} & $-10/\sqrt{168}$ & $-2/\sqrt{168}$ & $-1/\sqrt{168}$ \\ 
 \hline\hline
\end{tabular}
\end{center}
\label{table:1}
\end{table}
Since $\bar g_{i}$ are the canonically normalized couplings, their scale evolution at 1-loop is   
\begin{eqnarray}
  \frac{1}{\alpha_i(\mu)}=\frac{1}{\alpha_u}+\frac{1}{2 \pi} b_i \ln
  \left ( \frac{M_u}{\mu} \right),
\end{eqnarray}
where $\alpha_1=5/3 \bar g^2_1/(4\pi)=5/(3 \cos^2\theta_W) \alpha$, $\alpha_2=\bar g^2_2/(4\pi)$ and $\alpha_3=\bar g^2_3/(4\pi)$. Note that the factor $5/3$ appearing in the definition of $\alpha_1$ is the embedding coefficient of $U(1)$ into $SU(5)$.
The parameters $b_i$ are given by $b^{\rm{SM}}
_i=(b^{\rm{SM}}_1, b^{\rm{SM}}_2,
b^{\rm{SM}}_3)=(41/10, -19/6, -7)$ below the supersymmetric scale and by
$b^{\rm{S}}_i=(b^{\rm{S}}_1, b^{\rm{S}}_2, b^{\rm{S}}_3)=(33/5, 1,
-3)$ when ${\cal N}=1$ supersymmetry is restored.

Let us first show how to recover the results of  \cite{CF2002} within this new formalism. The case considered in \cite{CF2002} corresponds to $r=\bf{1}$ in which the coefficients are equal for all subgroups and $\epsilon_1 = \epsilon_2 = \epsilon_3$ in which case we have at the unification scale $M_u$:
\begin{equation}
\alpha_i(M_u)=\alpha_u(t,M_u).
\end{equation}

We thus have a mechanism to generate an effective time dependence of the unified coupling constant. Note that the unification condition is time independent and not affected by the dimension-five operators in Eq.~\eqref{L2}.

Note that the coupling constants $\alpha_i$ depend not only on the
scale $\mu$, but also on the time-dependent scalar fields $H_r$.
Since the coefficients $b_i$ are time independent, one finds
\begin{eqnarray} 
\label{renorminv}
\frac{1}{\alpha_i(\mu)}  \frac{\dot\alpha_i(\mu)}{\alpha_i(\mu)}
=
\frac{1}{\alpha_i(\mu')}  \frac{\dot\alpha_i(\mu')}{\alpha_i(\mu')}  \ \ \ i \in \{1,2,3\},
\end{eqnarray}
i.e., the quantity $\alpha_i^{-1} (\dot \alpha_i/\alpha_i)$ is scale
independent\footnote{By a dot we denote the time-derivative of the coupling constant. Thus, $\dot \alpha$ denotes the instantaneous time change of the fine-structure constant. The capital Greek letter $\Delta$, while not used in this paper, represents the difference in value of the coupling constant at different times, e.g., $\Delta \alpha=\alpha(t_2)-\alpha(t_1)$ with $t_2>t_1$. Finally, we will use the symbol $\delta$ to denote the time-dependent part of the coupling constant. For example, $\delta \alpha$ denotes the time-dependent part of the function $\alpha(t)$. Note that, sadly, authors in this field do not often define their notations carefully which may be the source of some confusion.}. 

We shall now calculate the relation between the time dependence of the fine structure constant $\alpha$ and the quantum chromodynamics scale $\Lambda_{\rm QCD}$. Following the same arguments as those presented in \cite{CF2002}, we obtain a relation between the instantaneous time changes of $\alpha$ and $\Lambda_{\rm QCD}$ given by
\begin{eqnarray} \label{result} 
\frac{\dot{\Lambda}_{\rm QCD}}{\Lambda_{\rm QCD}} = R \frac{\dot\alpha}{\alpha}(\mu=0).
\end{eqnarray}
The quantity $R$ is calculated in \cite{CF2002} and found to take the value $R=37.7\pm 2.3$. The uncertainty is due to low-energy non-perturbative quantum chromodynamics uncertainties. A time shift of the scale of quantum chromodynamics implies a time shift of the nucleon mass since its mass is proportional to that scale and determined to 90$\%$ by gluonic dynamics. Grand unification could simply be probed if time changes of the nucleon mass and the fine-structure constant were observed. 

Note that the relation (\ref{result}) (and thus $R$) is model independent in the sense that it does not depend on the scale at which supersymmetry is broken or the unification group. However, as we shall show shortly it is dependent on the choice of representation for the scalar multiplet.

In order to derive a relation of the type \eqref{result} for a generic representation, it is easiest to work out the coupling of $H_r$ to the low energy degrees of freedom from the standard model Lagrangian. 
The effective Lagrangian (\ref{L2}) is defined at the unification scale $M_u$. We can run it down to a low-energy scale $\mu$ and obtain
\begin{equation}
{\cal L} = -\frac{1}{4}\sum_{i=1}^{3}
\left( \frac{1}{g_u^2}+\sum_r \frac{c_r}{M_{\rm P}}H_r\delta_i^{(r)} + 
\frac{b_i}{8\pi^2} \log\left(\frac{M_u}{\mu}\right)\right)
 F_{\mu\nu (i)}{F}^{\mu\nu}_{(i)}.
\end{equation}
Below the supersymmetry breaking scale $M_{\rm S}$, one has
\begin{equation}
{\cal L} = -\frac{1}{4} \sum_{i=1}^{3}
\left( \frac{1}{g_u^2}+\sum_r \frac{c_r}{M_{\rm P
}}H_r\delta_i^{(r)} + \frac{b^{\rm{S}}_i}{8\pi^2}   \log\left(\frac{M_u}{M_{\rm S}}\right)+ \frac{b^{\rm{SM}}_i}{8\pi^2} \log\left(\frac{M_{\rm S}}{\mu}\right)\right)
 F_{\mu\nu (i)}{F}^{\mu\nu}_{(i)}.
\end{equation}
As we are working to leading order in perturbation theory this is equivalent to
\begin{equation}
{\cal L} = -\frac{1}{4}\sum_{i=1}^{3}
\left( 1+ g_i^2(\mu) \sum_r \frac{c_r}{M_{\rm P}}H_r\delta_i^{(r)}\right)
 F_{\mu\nu (i)}{F}^{\mu\nu}_{(i)}.
\end{equation}

Below the electroweak breaking energy scale, we can reformulate the effective Lagrangian in terms of the mass eigenstates: the photon $F_{\mu\nu}$ and the three massive electroweak bosons $Z_{\mu\nu}, W_{\mu\nu}$. In terms of the mass eigenstates and the $\epsilon_i$, the effective Lagrangian reads 
\begin{eqnarray} \label{L5}
{\cal L} &=& -\frac{1}{4}
 \left( 1+ e^2(\mu)(5\epsilon_1/3 + \epsilon_2)\right)
 F_{\mu\nu}{F}^{\mu\nu} \\ \nonumber &&
- \frac{1}{4}
 \left( 1+ e^2(\mu)\left(5\epsilon_1/3\tan^2\theta_W(\mu)+\epsilon_2\cot^2\theta_W(\mu) \right)\right) Z_{\mu\nu}{Z}^{\mu\nu} 
\\ \nonumber &&
-\frac{1}{2}\left( 1+ g_2^2(\mu) \epsilon_2\right)
{\rm Tr} W^{\dagger}_{\mu\nu}W^{\mu\nu}
-\frac{1}{2}\left( 1+ g_3^2(\mu)\epsilon_3\right){\rm Tr}
 G_{\mu\nu}{G}^{\mu\nu},
\end{eqnarray}
where the factor $5/3$ in the first equation is the embedding coefficient of $U(1)$ into $SU(5)$. As expected, the photon and $Z$ boson interactions show mixing between $U(1)_Y$ and $SU(2)_L$ couplings $\epsilon_1, \epsilon_2$. Similarly, the $W$ boson and gluon terms are exclusively associated with $\epsilon_2$ and $\epsilon_3$, respectively. Modulo the Clebsch-Gordon coefficients $\delta_i^{(r)}$, and as a consequence of the embedding in a unified theory, all modifications are proportional to the combination $\sum_r (c_r/M_{\rm P})H_r$ and the coupling-constant variations are interdependent.

From Eqs.~\eqref{timedepcou},\eqref{L5}, we identify the fine-structure constant $\alpha$ and strong coupling constant $\alpha_s$:
\begin{eqnarray}
\alpha(\mu) &=& \frac{\bar e^2(\mu)}{4\pi} \approx\frac{e^2(\mu)(1-e^2(5\epsilon_1/3 + \epsilon_2))}{4\pi}, \label{alphamu}\\ \alpha_s(\mu) &=& \frac{\bar g_3^2(\mu)}{4\pi}\approx\frac{g_3^2(\mu)(1-g_3^2(\mu)\epsilon_3)}{4\pi},
\end{eqnarray}
which gives the instantaneous time changes
\begin{eqnarray}
\frac{\dot\alpha(\mu)}{\alpha(\mu)} \approx -4 \pi \alpha(\mu)(5\dot\epsilon_1/3 + \dot\epsilon_2) , \qquad \frac{\dot\alpha_s(\mu)}{\alpha_s(\mu)} \approx -4 \pi \alpha_s(\mu)\dot \epsilon_3.
\label{newalphadot}
\end{eqnarray}
Similarly, the instantaneous time change of $\Lambda_{\rm QCD}$ is given by
\begin{eqnarray}
\label{QCDdot}
\frac{\dot\Lambda_{\rm QCD}}{\Lambda_{\rm QCD}}&=&\frac{1}{2}\log\left(\frac{\mu^2}{\Lambda_{\rm QCD}^2}\right)\frac{\dot\alpha_s(\mu)}{\alpha_s(\mu)} \approx \frac{8\pi^2}{b_3}\dot\epsilon_3,
\end{eqnarray}
where $b_3 = -11 + (2/3)n_f$ for $n_f$ quark flavors. Equations~\eqref{newalphadot} and~\eqref{QCDdot} 
express the time changes of $\alpha$, $\alpha_s$, and $\Lambda_{\rm QCD}$ in terms of the time changes of the scalar fields. 

Using ~\eqref{newalphadot} and~\eqref{QCDdot}, we obtain:
\begin{eqnarray}
\label{QCDdotgen}
\frac{\dot\Lambda_{\rm QCD}}{\Lambda_{\rm QCD}}&\approx& R_r
\frac{\dot\alpha}{\alpha}(\mu=0)
\end{eqnarray}
where
{\begin{eqnarray}
R_r= -\frac{2\pi}{b_3}\frac{1}{\alpha} \
\frac{\sum_r c_r\dot H_r\delta_3^{(r)}}{(5/3)\sum_r c_r \dot H_r\delta_1^{(r)}+\sum_r c_r \dot H_r\delta_2^{(r)}}.
\end{eqnarray}}
This relation is obtained to leading order as expansions in $1/M_P$. Because of this expansion, this relation is not renormalization group invariant. However, it enables us to see that depending on the choice of representation we can obtain different predictions for $R_r$.  For $r=1$, we find $R_1=46$ where the discrepancy with respect to \eqref{result} comes from the fact that this latter approach is not renormalization group invariant. For $r=24$, we find $R_{24}=-53$, for $r=75$, we find $R_{75}=-23$ and for $r=200$, we find $R_{200}=7$. Note that we used $b_3=-7$ for these estimates. Clearly the quantity $R_r$ is sensitive to the number of scalar multiplets and their representations. It can be negative or positive. 

In \cite{CF2002}, it was shown that a negative sign for $R_r$ could also be due to a time variation of the scale of unification. Discriminating between different models of time variation using only low energy experiments may thus prove difficult. It is thus going to be important to measure the couplings of the different scalar fields responsible for a time variation to standard model particles. This is the question we address in the next section in the context of ultra light dark matter.

\section{Ultra light dark matter and constraints}
Thus far, we have treated the scalar multiplets $H^{ab}_r$ as generic time-dependent classical fields. Let us now consider the case where $H_r$ are ultra light and represent dark matter. 
At energies $\mu \lesssim 1$\;GeV of atomic-scale experiments, electrons, light quarks, and their associated gauge fields are the most relevant degrees of freedom. From this point of view, the interesting part of the effective Lagrangian Eq.~\eqref{L5} are the first and last terms parametrizing the interactions of four ultra light dark-matter candidates to photons and gluons.
Given this assumption, the functional form $H_r(t)$ is well approximated by a single-mode solution to the Klein-Gordon equation~\cite{DP14,SF14,AHT15}
\begin{eqnarray}
\label{HDM}
H_r(t)\approx \frac{\sqrt{2 \rho_{H_r}}}{m_{H_r}} \cos(m_{H_r} t),
\end{eqnarray}
where $m_{H_r}$ is the mass and  $\rho_{H_r}$ is some fraction of the dark-matter density ($\rho_{\rm DM} \approx 0.4$\;GeV/cm$^3$). This solution generates periodic oscillations of the fine-structure constant $\alpha$ and strong coupling constant $\alpha_s$. 

Searches for ultra light dark matter have been performed in a variety of experiments~\cite{SMDM}. As applications, we consider three types of experiments that together yield the most stringent constraints on the coefficients $c_r$ in the mass range $10^{-24}$\;eV $\lesssim m_{H_r} \lesssim 10^{-14}$\;eV. These include optical atomic clocks, in particular those of the Physikalisch-Technische Bundesanstalt (PTB)~\cite{Filzinger23}; pulsar timing array observations from the NANOGrav Collaboration~\cite{KMT22,NANOGrav23}; and weak equivalence principle tests by the MICROSCOPE Collaboration~\cite{Berge:2017ovy,MICROSCOPE:2022doy}. 

Optical atomic clock experiments measure the ratio $f_A/f_B$ of two transitions $A$ and $B$. The time-dependent part of this ratio is typically expressed as a relative fraction and parametrized in terms of the relative time-dependent contributions of the fine-structure constant $\alpha$, electron mass to QCD scale $m_e/\Lambda_{\rm QCD}$, and light-quark masses to QCD scale $m_q/\Lambda_{\rm QCD}$ as~\cite{Flambaum04}
\begin{align}
\label{varratio}
\frac{\delta (f_A/f_B)}{f_A/f_B} &= (k_A^\alpha - k_B^\alpha) \frac{\delta \alpha}{\alpha} + (k_A^\mu - k_B^\mu)\frac{\delta(m_e/\Lambda_{\rm QCD})}{m_e/\Lambda_{\rm QCD}} + (k_A^q - k_B^q)\frac{\delta(m_q/\Lambda_{\rm QCD})}{m_q/\Lambda_{\rm QCD}} \nonumber\\
& = (k_A^\alpha - k_B^\alpha) \frac{\delta \alpha}{\alpha} - (k_A^\mu - k_B^\mu + k_A^q - k_B^q)\frac{\delta\Lambda_{\rm QCD}}{\Lambda_{\rm QCD}},
\end{align}
where the proportionality factors $k_A^\alpha, k_A^\mu$, and $k_A^q$ quantify the sensitivity of the transition $A$ to the associated dimensionless quantity. Note that since the effective Lagrangian Eq.~\eqref{L5} contains only gauge-sector modifications by assumption\footnote{Note that in realistic models, one expects the time variation of lepton and quark masses to be subdominant effects \cite{CF2002}.}, there is no time dependence of the electron $m_e$ and quark $m_q$ masses ($\delta m_e = \delta m_q = 0$). 
From Eq.~\eqref{L5}, we extract the relative time-dependent pieces\footnote{With these definitions, e.g., $\dot\alpha = d(\delta\alpha)/dt$.}
\begin{equation}
\label{varQEDQCD}
\frac{\delta \alpha}{\alpha} \approx -4 \pi \alpha(\mu)(5\epsilon_1/3 + \epsilon_2), \qquad
\frac{\delta \alpha_s}{\alpha_s} \approx-4\pi\alpha_s\epsilon_3, \qquad \frac{\delta\Lambda_{\rm QCD}}{\Lambda_{\rm QCD}} \approx \frac{8\pi^2}{b_3}\epsilon_3.
\end{equation}
Substituting these with Eq.~\eqref{HDM} into Eq.~\eqref{varratio}
gives 
\begin{equation}
\label{clocksignal}
\frac{\delta (f_A/f_B)}{f_A/f_B} 
=\frac{1}{M_{\rm P}}\sum_r( k^r_A-k^r_B) c_r\frac{\sqrt{2 \rho_{H_r}}}{m_{H_r}} \cos(m_{H_r} t),
\end{equation}
where
\begin{align}
k^r_A-k^r_B \equiv (k_A^\alpha - k_B^\alpha)\left(-4 \pi \alpha(\mu)(5/3\delta_1^{(r)} + \delta_2^{(r)})\right)  - (k_A^\mu - k_B^\mu + k_A^q - k_B^q)(8\pi^2/b_3)\delta_3^{(r)}.
\end{align}

If present, the signal Eq.~\eqref{clocksignal} would appear as a single peak, or multiple peaks, in the Fourier amplitude spectrum at angular frequencies $\omega \approx m_{H_r}$. PTB have measured the ${}^{171}$Yb$^+$ electric-octupole (E3) transition relative to electric-quadrupole (E2) transition ($f_{\rm E3}/f_{\rm E2}$) and the E3 transition relative to the ${}^{87}$Sr 
(${}^1$S$_0 \leftrightarrow {}^3$P$_0$) transition ($f_{\rm E3}/f_{\rm Sr})$ over approximately 2 years~\cite{Filzinger23},\cite{AxionLimits}. These optical-to-optical comparisons are primarily sensitive\footnote{It has been suggested that optical-clock comparisons also admit sensitivity to nuclear radius oscillations, which could be sourced by the second term in Eq.~\eqref{dilaton}~\cite{Banerjee:2023bjc}. However, the associated sensitivity factor is estimated to be suppressed relative to $k^\alpha$ by $\approx 10^3$, resulting in significantly weaker constraints on $d_g$.} time-dependent contributions $\delta \alpha/\alpha$ from Eq.~\eqref{varratio}. The non-observation of significant peaks in the amplitude spectrum was translated into constraints on $|d_\gamma|$ from 
single dilaton-like field $\phi$ model~\cite{DD2010} 
\begin{equation}
\label{dilaton}
\mathcal{L}_{\phi} =  \frac{1}{\sqrt{2}M_{\rm P}}\phi\left[\frac{1}{4e^2}d_\gamma F_{\mu\nu}F^{\mu\nu} -\frac{\beta_3}{g_3}d_g{\rm Tr}G_{\mu\nu}G^{\mu\nu}\right],
\end{equation}
in the range $10^{-23}$\;eV $\lesssim m_{\phi} \lesssim 10^{-17}$\;eV,
where $\beta_3 = -b_3 g_3^3/(16\pi^2)$ is the quantum chromodynamics beta function. In this model, $\delta \alpha/\alpha = d_\gamma\phi(t)/(\sqrt{2}M_{\rm P})$ with $\phi(t) = (\sqrt{2\rho_{\rm DM}}/m_\phi)\cos(m_\phi t)$ in analogy with Eq.~\eqref{HDM}.
Integrating out the $Z$ boson, $W$ boson, and heavy quarks from Eq.~\eqref{L5}, comparing at $\mu \approx 1$\;GeV, and assuming only one of the four irreducible representations $r$ is present, allows constraints on $d_\gamma$ and $d_g$ to be recast in terms of constraints on the coefficients $c_r$.
Comparing the fractional shift with those obtained from the dilaton model Eq.~\eqref{dilaton}, we find that, for a given irreducible representation $r$:
\begin{align}
\label{mapping}
\sqrt{2}\left(-4 \pi \alpha(\mu)(5/3\delta_1^{(r)} + \delta_2^{(r)})\right) c_r&\leftrightarrow d_\gamma, \nonumber\\ 
\sqrt{2}(8\pi^2/b_3)\delta_3^{(r)}c_r &\leftrightarrow d_g. 
\end{align}
Under the assumption that a single irreducible representation $r$ is present and accounts for the total dark-matter density ($\rho_{H_r} = \rho_{\rm DM})$, constraints on each of the $c_r$ may be extracted one at a time by setting the remaining three to zero.

Pulsar timing arrays provide sensitivity to ultra light dark matter through timing residuals $h(t)$, which quantify the difference between the time of arrival of a pulsar's measured electromagnetic signal relative to a model prediction~\cite{KMT22}. Cesium atomic clocks are used as references for pulsar time-of-arrival measurements and are sensitive to the time-dependent contributions $\delta f_{\rm Cs}/f_{\rm Cs}$ induced by ultra light dark matter through the time-dependent pieces of $\alpha$, $m_e/\Lambda_{\rm QCD}$, and  $m_q/\Lambda_{\rm QCD}$ from Eq.~\eqref{varratio}\footnote{Note that here one must also account for the spatial coherence properties of ultra light dark matter (see ~\cite{KMT22}).}. A change of the reference cesium clocks' rate would appear as a change in the pulsar rotation frequency $f$ (i.e. $\delta f_{\rm Cs}/f_{\rm Cs} = \delta f/f$), the latter of which is related to the timing residuals $h(t)$ via
\begin{equation}
\label{resid}
h(t) = -\int \frac{\delta f}{f}dt.
\end{equation}
Applications of the NANOGrav Collaboration 15-year data set include constraints on the ultra light dark matter dilaton coefficients $d_\gamma, d_g$ ~\cite{NANOGrav23}. The constraints on $d_g$ are roughly an order of magnitude more stringent than microwave atomic-clock measurements~\cite{Hees:2016gop} for masses $10^{-24}$\;eV $\lesssim m_{\phi} \lesssim 10^{-23}$\;eV. This translates to a dominant sensitivity to our $c_r$ coefficients through the identification Eq.~\eqref{mapping} over this same range. 

The presence of ultra light bosonic\footnote{In this case $H_r$ need not necessarily be identified with dark matter.} fields $H_r$ will also generate a fifth force, corresponding to a Yukawa potential between two objects $A, B$ of different composition. 
The MICROSCOPE Collaboration have obtained the most stringent limit on the E{\"o}tv{\"o}s parameter $\eta = 2|\vec{a}_A-\vec{a}_{B}|/|\vec{a}_A+\vec{a}_{B}| \approx 10^{-15}$ using titanium and platinum alloys~\cite{MICROSCOPE:2022doy}. Matching $\eta$ to a combination of Wilson coefficients of the effective photon and gluon operators in Eq.~\eqref{L5} and functions of the proton and mass numbers of $A$ and $B$~\cite{Berge:2017ovy,Brzeminski:2022sde,Hees:2018fpg} enables constraints on $c_r$. 
\begin{figure}[h!]
\centering
\adjustbox{center=\textwidth}{\includegraphics[width=0.8\textwidth,angle=-90]{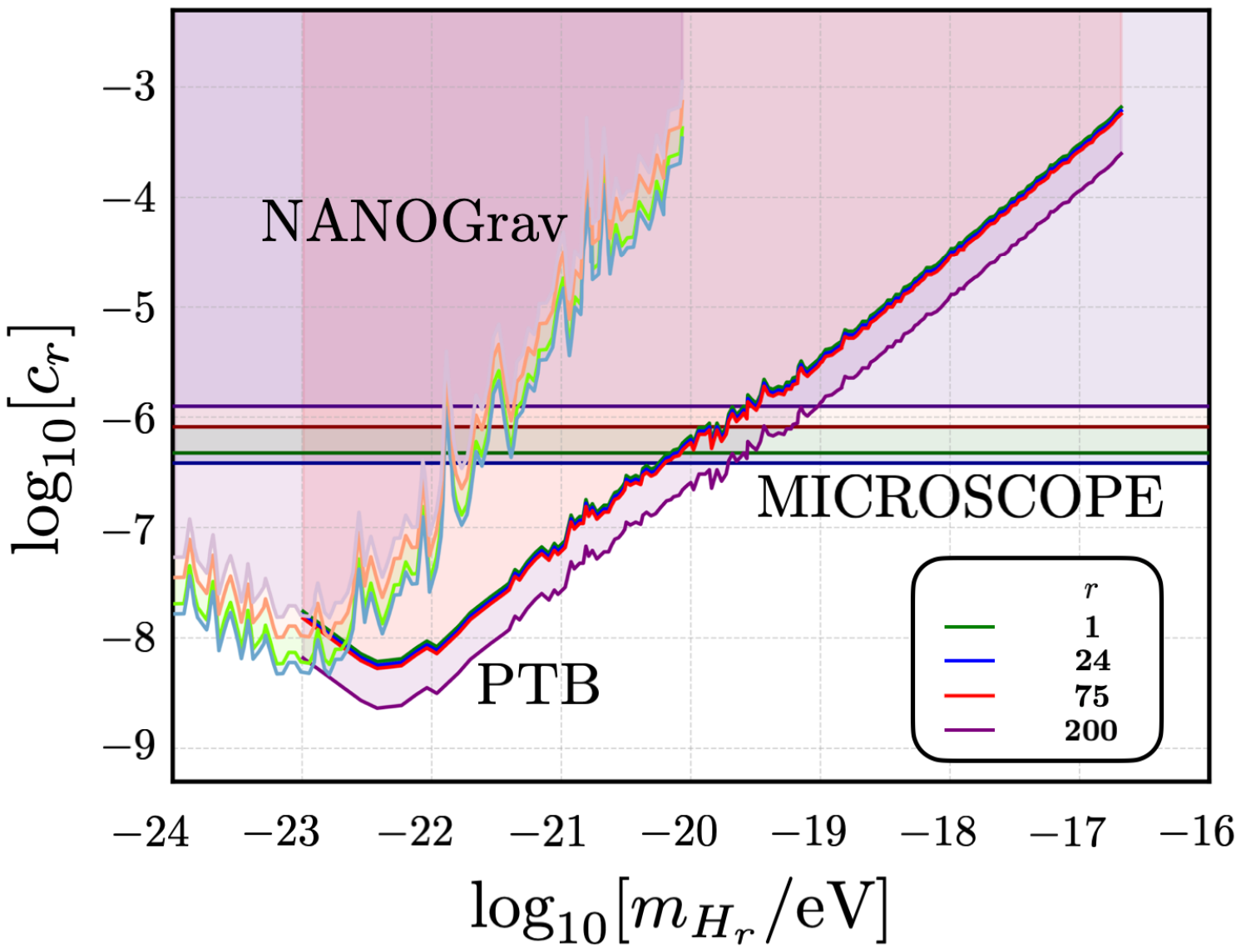}}
\vspace{-0.75cm}
\caption{Constraints on $c_r$. Multiplets $r = $ \textbf{1}, \textbf{24}, \textbf{75}, \textbf{200} are represented by green, blue, red, and purple lines, respectively. Lighter (darker) variations of the color palette apply to NANOGrav (MICROSCOPE).}
    \label{fig:constraints}
\end{figure}

Constraints at 95\% confidence level on $c_r$ are shown in
Fig.~\ref{fig:constraints}. We have assumed a renormalization scale $\mu$ corresponding to the typical scale of atomic physics ($\alpha\approx 1/137$) and $b_3 =-9$, the latter corresponding to $n_f = 3$ light-quark flavors (and subject to theory uncertainties from non-perturbative quantum chromodynamics---see Eq.~\eqref{result}). The small differences in $c_r$ exclusion regions are completely controlled by the Clebsch-Gordon coefficients $\delta_i^{(r)}$ listed in Table~\ref{table:1}. 
As mentioned, NANOGrav gives the most sensitivity in the lowest mass range $10^{-24}$\;eV $\lesssim m_{H_r} \lesssim 10^{-23}$\;eV and along with PTB exceed the MICROSCOPE constraints until around $m_{H_r}\approx 10^{-20}$\;eV. 

Note that MICROSCOPE has increased sensitivity to gluonic interactions relative to electromagnetic interactions (see, e.g., Ref.~\cite{Brzeminski:2022sde}). In our unified framework, both of these interactions naturally
appear and the former are typically dominant as indicated in Eq.~\eqref{mapping}. This explains the flat constraints on $c_r\sim 10^{-6}$. As a visual preference, we show the MICROSCOPE constraints through $m_{H_r} \approx 10^{-16}$\;eV though they hold at the same magnitudes until $m_{H_r} \approx 10^{-14}$\;eV before rapidly falling off when the wavelength of $H_r$ becomes comparable to the radius of the Earth. We also note that several astrophysical and cosmological observations indicate $m_{H_r} \gtrsim 10^{-21}$\;eV~\cite{Ferreira20} and that lighter dark-matter candidates can only account for $\sim 10\%$ of the total dark-matter density $\rho_{\rm DM}$, which is also largest source of theory uncertainty for the NANOGrav and PTB bounds. 

Grand unification can naturally accommodate multiple scalar dark matter candidates. In $SU(5)$ grand unified theories, each of the scalar representations $r=$ \textbf{1}, \textbf{24}, \textbf{75}, \textbf{200} contains one neutral scalar field that could be a good dark matter candidate. They could all contribute to the local dark matter density. In the (rather restrictive) case of equal masses, we note that, assuming $\sum_r \rho_{H_r} = \rho_{\rm DM}$, constraints may placed analogously to the single-field case on the linear combination $\sum_r (k^r_A - k^r_B) c_r \sqrt{2\rho_{H_r}}$.

\section{Conclusions}

In this paper, we have shown that scalar multiplets in grand unified theories can naturally lead to a time-variation of fundamental constants via dimension five operators expected to be generated by quantum gravity.
These scalar multiplets naturally provide a mechanism for a time variation of the gauge coupling of the unified theory. We illustrate the mechanism using supersymmetric $SU(5)$ grand unified theory, but our conclusions would apply to any grand unified gauge theory. We emphasize that the unification of the gauge couplings of the standard model can be probed at very low energy with quantum sensors and other tabletop experiments. Grand unification imposes a relation between the time variation of the fine-structure constant and that of the scale of quantum chromodynamics or equivalently the proton mass. Our results do not depend on the details of the unification group or on the scale at which supersymmetry is broken. 

We then propose to identify the new scalar multiplets with ultra light dark matter. This assumption implies a specific functional dependence on time for the scalar multiplets. We use data from PTB, NANOGrav and MICROSCOPE to set a limit on the coefficient of the dimension five operators generated by quantum gravity. 

	\bigskip
	
	{\it Acknowledgments:}
We would like to thank an anonymous referee for their valuable questions which have helped us to improve our paper. This work is supported in part by the Science and Technology Facilities Council (grant numbers ST/T006048/1 and ST/Y004418/1). 	\\

	\bigskip 
	
	{\it Data Availability Statement:}
	This manuscript has no associated data. Data sharing not applicable to this article as no datasets were generated or analysed during the current study. 
	
	\bigskip 
	
	%\newpage

\end{document}